# A calculation method to estimate thermal conductivity of high entropy ceramic for thermal barrier coatings

Yuxuan Wang[a], Guoqiang Lan[a], Jun Song[a,*]

a Department of Mining and Materials Engineering, McGill, University, Montreal, Quebec, H3A 0C5, Canada

* Corresponding author. E-mail address: jun.song2@mcgill.ca (J. Song).

## Abstract

High entropy ceramics are highly promising as next generation thermal barrier coatings due to their unique disorder structure, which imparts ultra-low thermal conductivity and good high temperature stability. Unlike traditional ceramic materials, the thermal resistance in high entropy ceramics predominantly arises from phonon-disorder scattering rather than phonon-phonon interactions. In this study, we propose a calculation method based on the supercell phonon unfolding (SPU) technique to predict the thermal conductivity of high entropy ceramics, specially focusing on rocksalt oxides structures. Our prediction method relies on using the reciprocal value of SPU phonon spectra linewidth as an indicator of phonon lifetime. The obtained results demonstrate a strong agreement between the predicted thermal conductivities and the experimental measurements, validating the feasibility of our calculation method. Furthermore, we extensively investigate and discuss the atomic relaxation and lattice distortion effects in 5-dopants and 6-dopants rocksalt structures during the process.

## 1. Introduction

Thermal barrier coating materials have been extensively used in fields of aerospace and power industry[1-6] to resist high temperature environment. Superalloys used in turbine blades typically have a melting point of approximately 1300 °C, making them unsuitable for direct exposure to hotter gas streams. Under this circumstance, thermal barrier coating plays a crucial role, which can protect the underlying substrate against corrosive hot gas streams while effectively isolating the heat flow, thereby reducing the temperature reaching to the metal surface and enabling the metal components to operate at temperatures exceeding their melting points. Currently, the primary challenge in the development of thermal barrier coating materials revolves around their suitability for higher service temperature as increased temperature usually means enhanced power-to-weight ratios, improved fuel burning efficiency, and reduced by-product waste emissions, all of which have led to an increasing demand for high-performance thermal barrier coating materials.

Thermal barrier coating (TBCs) generally refers to the outermost ceramic isolation layer, and two significant performance index of which are low thermal conductivity and high temperature phase stability[3, 7, 8]. Low thermal conductivity ensures a significant temperature drop from the coating surface to the metal component, while high temperature thermal stability guarantees the integrity of the coating in a high temperature environment. Currently most-used commercial TBC materials is 7~8 wt% $Y_2O_3$ stabilized zirconia (7-8YSZ). However, when temperature keeps higher than 1200 °C, the t'-tetragonal phase will transform to another tetragonal phase (t) and/or a cubic phase (c) and then to monoclinic (m) phase [9, 10]. This phase transition compromises the integrity of the coating. To address this issue and develop the next generation TBC materials, extensive researches have been conducted on complex oxides such as pyrochlore oxides [11-18] and silicate oxides[19-22]. And recently, a new class of complex oxides - high entropy oxides have garnered considerable attention. The concept of high entropy originates from alloys. Alloys containing five or more elements in relatively high concentrations (5-35 at%) or having configurational entropy larger than 1.5R belong to high entropy alloys[23]. Being broken on translational symmetry imparts high entropy materials some unique properties, including prominent strength, toughness, low electrical and thermal conductivity. Besides, according to Gibbs free energy expression $G = H - TS$, higher entropy can facilitate the drop of free energy at high temperature and hence encourage the stabilization of phase structure[24]. Researchers have started applying the concept of high entropy to traditional low-thermal-conductivity structures such as fluorite oxides[25-28] and pyrochlore oxides[29-33] and obtained substantial experimental results. However, unlike traditional one or two dopants structure, high entropy oxides typically contains five or more dopant elements and have plenty of permutation combination, poses a challenge for experimental screening, not to mention the potential material waste involved in the process.

The development of high-performance computer cluster makes computational materials screening possible, greatly simplifying the tedious experimental process. However, the conventional calculational model employed for predicting thermal conductivity encounters limitation when applied to high entropy materials. On one hand, strictly, high entropy oxides lack crystallographic symmetry, which means no primitive cell can be identified for subsequent analysis; On the other hand, for general ceramic materials, thermal conductivity mainly come from phonon-phonon interaction, which can be evaluated by solving the Boltzmann Transport Equation (BTE)[34]. But for high entropy oxide, phonons interaction will be complicatedly influenced by mass disorder and force disorder. To address the crystallographic problem, although does not follow strict symmetry, long-term order can still be detected experimentally in high entropy materials. It means if ignore the force and mass differences and lattice distortion on dopants position, minimal periodic unit can be defined. Therefore, some mean-field theories, like Virtual Crystal Approximation (VCA)[35] and (coherent potential approximation) CPA[36, 37], are proposed to deal with the mass and force disorder. However, these approaches neglect the local differences of atomic interaction, thereby failing to accurately reflect the intricate nature in high entropy oxides. More recently, supercell phonon unfolding (SPU)[38, 39] method has emerged as a promising alternative. This method considers the interaction from every individual atomic pairs and reflect it on phonon dispersion spectra, which makes it more precise for describing the disordered system. It maps the phonon dispersion of supercell to that of primitive cell, at the same time keeping all the information, and thus make it easier to compare the result with that of similar primitive structure. Moreover, several studies have shown that the phonon dispersion spectra offer an accurate portrayal of vibrational physics, with the linewidth of the spectra inversely proportional to the phonon lifetime[37, 40, 41]. This finding afford an opportunity to construct thermal conductivity model specially tailored for high entropy ceramic materials.

This study aims to construct the supercell structure for high entropy rocksalt oxides and utilized the supercell phonon unfolding (SPU) method to calculate the phonon dispersion spectra. And based on these calculation, a prediction model for thermal conductivity is developed. The composition design of high entropy rocksalt oxides derived from Braun's research[42], encompassing one 5-dopant oxide and five 6-dopant oxides, where experimental results for thermal conductivity are available for comparison with the predicted values. Additionally, the atomic relaxation and lattice distortion of these six high entropy rocksalt oxides are thoroughly investigated and analyzed in this study.

## 2. Methodology

2.1 Density functional theory and supercell construction

Supercell structures optimization and relative atomic force constants were calculated by DFT (density functional theory), performed using Vienna ab initio simulation package (VASP). For pseudopotential treatment, Perdew-Burke- Ernzerhof (PBE) GGA approach based on plane-wave base sets was adopted. Valence configuration in this study of different elements are as follow: Mg - $2p^6\,3s^2$, Zn - $3d^{10}\,4s^2$, Co - $3d^8\,4s^1$, Cu - $3p^6\,3d^{10}\,4s^1$, Ni - $3p^6\,3d^9\,4s^1$, Sc - $3s^2\,3p^6\,3d^1\,4s^2$, Sb - $5s^2\,5p^3$, Sn - $4d^{10}\,5s^2\,5p^2$, Cr - $3p^6\,3d^5\,4s^1$, Ge - $3d^{10}\,4s^2\,4p^2$ and O - $2s^2\,2p^4$. The value of Hubbard energy for $3d$ orbit of Cr, Co and Ni were set following Jain's research[43]. The electron - core interaction was described by the Bloch's projector augmented wave method (PAW) within the frozen-core approximation.

The supercell construction was realized by SQS (special quasi-random supercells) method through Python package *ICET*. Only the positions of metal atoms were randomly replaced by different equiatomic dopants while keeping unchanged of oxygen positions. For 5-dopant oxide $(Mg_xNi_xCu_xCo_xZn_x)O$ (J14, $x = 1/5$), primitive cell was used to extend to its 5×5×5 supercell (250 atoms) while conventional cell was used to extend to its 3×3×3 supercell (216 atoms) for other 6-dopant oxides $(Mg_xNi_xCu_xCo_xZn_xX_x)O$ (J30, J31, J34, J35, J36 for X = Sc, Sb, Sn, Cr, Ge, $x = 1/6$) to ensure the number of every element to be integers. For every composition, 5 random structures were created for repeated trials. And for structure optimization, cell shape and atomic positions were firstly fixed and the lattice constants with lowest energy were calculated. And then lattice constants were fixed to relax the atom positions. Gamma-centered k-mesh of $1 \times 1 \times 1$ for Brillouin-zone integrations and cutoff energy of 520 eV for the plane-wave-basis were used in DFT calculation with convergence criteria for energy on each ion set as $5 \times 10^{-8}$ eV. The real-space force constants were calculated with finite displacement method using $1 \times 1 \times 1$ supercell. And phonon properties were evaluated using PHONOPY package on $20 \times 20 \times 20$ q grid for all the high entropy oxide samples.

## 2.2 SPU and thermal conductivity indicator

Supercell phonon-unfolding method stems from a kind of band-unfolding treatment dealing with electron and phonons in random alloying system[39]. By unfolding the band of large random alloy supercell onto that of small primitive-cell periodic basis, an effective band spectra can be obtained. In this paper, we used the unfolding program *upho* written by Ikeda etc[40]. The unfolded spectral function is defined as: $A(\boldsymbol{k}_k, \omega) = \sum_J \left|\left[\hat{P}^{\boldsymbol{k}_k}\tilde{v}(\boldsymbol{K},J)\right]_l\right|^2 \delta[\omega - \omega(\boldsymbol{K},J)]$, where $\boldsymbol{k}_k$ is wave vector of primitive cell while $\boldsymbol{K}$ is from supercell, $J$ is band index and $\delta$ means delta function. In this expression, $\hat{P}^{\boldsymbol{k}_k}$ represents projection operator at $\boldsymbol{k}$ point in the primitive Brillouin zone, and $\tilde{v}(\boldsymbol{K},J)$ denotes the eigenvectors of dynamical matrix of the supercell for wave vector $\boldsymbol{K}$. Delta function $\delta$ were smeared by the Lorentzian functions with half width at half maximum (HWHM) of 0.05THz for spectra plotting. And the linewidths of phonon band were defined as the full width at half maximum (FWHM) of unfolded spectra through Lorentzian function fitting. Referring Boltzmann equation with a relaxation time approximation to describe thermal conductivity[44], we built a thermal conductivity prediction indicator $\Lambda$ of high entropy ceramics. Consider that phonon linewidth is inversely proportional to phonon lifetime, indicator $\Lambda$ can be written as: $\Lambda = \frac{A}{NV_0} \sum_\lambda \frac{Cv_\lambda \cdot v_{\lambda,x}^2}{lw_\lambda}$, where $A$ is adjustment parameter, $V_0$ is volume of primitive cell, $N$ is number of wave vector, $C_v$ is heat capacity, $v$ is phonon velocity, and $lw$ is phonon linewidth.

# 3. Results and discussion

## 3.1 Structure optimization

The space group symbol of rocksalt oxide structure is *Fm3m*, in which oxygen atom occupies the 4b position while metal atom takes the 4a position. Fig. 1 illustrates the supercell structure of SQS constructed high entropy oxide. Fig.1 (a) is parallelepiped supercell and belongs to 5-dopant high entropy oxide J14. The small red balls are oxygen atoms and other colored balls are different metal dopants. The total number of metal atoms is 125, and every dopant element has 25 atoms. Fig.1 (b) is cubic supercell structure and belongs to other five 6-dopant oxide J30, J31, J34, J35, J36. The total number of metal atoms is 108 and for every dopant element they has 18 atoms. SQS is the currently considered the most reliable modeling method for studying high entropy system. It allows for the random distribution of different types of atoms within the structure. In some experimental results, short-range ordering structures may also be observed in high entropy materials[45, 46]. Within this paper, we assume that the high entropy effect dominates the mechanism of phase stabilization and ignore the minute effect of local ordering.

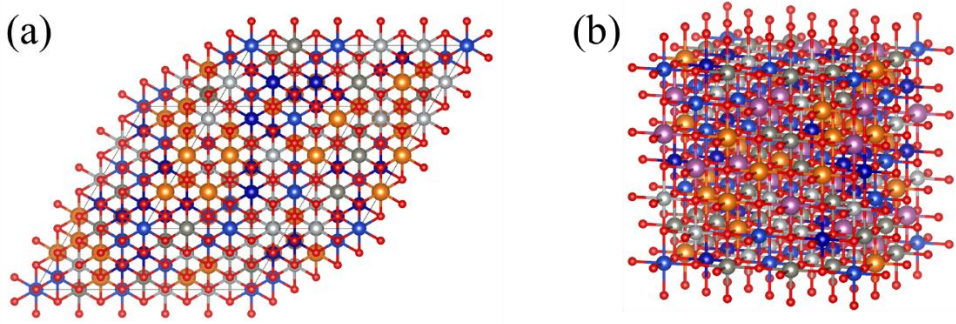

Figure 1 Supercell construction of (a) J14 and (b) other 6-dopant high entropy oxides.

Table 1 presents the compositions, calculated lattice constants and experimental thermal conductivities of 6 kind of rocksalt high entropy oxides. It is evident that the measured thermal conductivities are nearly halved after the addition of the sixth dopant element. Except for J14, the thermal conductivities of the other 6-dopant high-entropy oxides exhibit a negative correlation with lattice constants. The J31 has the highest lattice constant 3.154 Å, which corresponds to the lowest thermal conductivity 1.41 $W \cdot m^{-1} \cdot K^{-1}$ while J30 owns the lowest lattice constant 3.030 Å but with highest thermal conductivity 1.68 $W \cdot m^{-1} \cdot K^{-1}$. This relationship reminds of Slack's model of thermal conductivity, which suggests that the average volume of atom is proportional to thermal conductivities. The influencing factors of Slack's model also comprise the average mass and Gruneisen's parameter, which denotes the nonlinearities of the phonons. However, when consider the case of J14 in, this relationship becomes unclear. The average masses of these compositions are quite similar, and the addition of the sixth element does not significantly alter the Gruneisen's parameter. Therefore, there must be other mechanisms contributing to this phenomenon.

Table 1 Compositions, calculated lattice constants and experimentally obtained thermal conductivities.

| Name | Composition | a(Å) | Ex. Thermal conductivity ($W \cdot m^{-1} \cdot K^{-1}$)[42] |
|---|---|---|---|
| J14 | $(Mg_xNi_xCu_xCo_xZn_x)O$, $x=0.2$ | 3.020 | 2.95 ± 0.49 |
| J30 | $(Mg_xNi_xCu_xCo_xZn_xSc_x)O$, $x=0.167$ | 3.030 | 1.68 ± 0.13 |
| J31 | $(Mg_xNi_xCu_xCo_xZn_xSb_x)O$, $x=0.167$ | 3.154 | 1.41 ± 0.17 |
| J34 | $(Mg_xNi_xCu_xCo_xZn_xSn_x)O$, $x=0.167$ | 3.127 | 1.44 ± 0.10 |
| J35 | $(Mg_xNi_xCu_xCo_xZn_xCr_x)O$, $x=0.167$ | 3.035 | 1.64 ± 0.24 |
| J36 | $(Mg_xNi_xCu_xCo_xZn_xGe_x)O$, $x=0.167$ | 3.075 | 1.60 ± 0.14 |

Following the volume optimization, atomic relaxation were conducted and atomic displacements from their ideal positions are presented in Figure 2. It's evident that the displacements distribution of J14 differs from other compositions. In the case of J14, the displacements of all the elements are smaller than 0.5 angstrom, and they remain close to their ideal positions. Notably, the magnitude of displacement for oxygen atoms is nearly twice as large as that for the other metal elements. An intriguing phenomenon is that this distribution of atomic displacement changed a lot after the addition of 6$^{th}$ dopant element. In particular, the farthest position change of copper atoms reaches about 1.5 angstrom, which is almost half of the primitive lattice vector. For J30 and J35, their atomic displacement distribution are quite similar and their thermal conductivities are also quite close to each other. Except for the wide distribution of copper atoms, the distribution of other elements keep in a relatively narrow range. And for the cases of J31, J34, their atomic displacement and distributions are distinctly different from 5-dopant oxide J14. Besides the wider distribution of copper atoms, other elements also exhibit a broad range of displacement, which indicate their severe lattice distortion, corresponding to their low thermal conductivity. And for J36, its distribution lies somewhere between J30 and J31. From this comparation, it appears that the substantial change of copper atom displacement distribution may be responsible for the abruptly reduction of thermal conductivity after the addition of the sixth dopant. Further analysis of the oxygen environment of each metal-oxygen polyhedron will be discussed in the next subchapter.

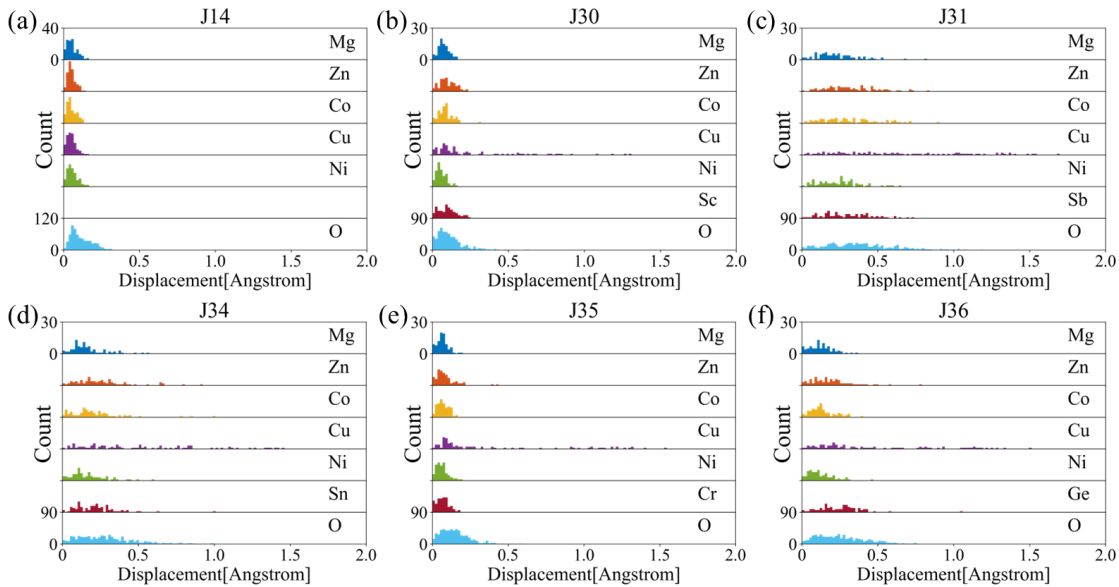

Figure 2 Atomic displacement distribution of rock salt high entropy oxides.

## 3.2 Lattice distortion

Lattice distortion is an important indicator of many properties of high entropy materials[47-51]. To deeply investigate the lattice distortion, we analyzed the 1st nearest neighbor (1NN) metal-oxygen coordinating distance in [MO$_6$] polyhedrons. Fig. 3 is the histogram of 1NN M-O distance distribution. For the J14 composition, each type of metal-oxygen bonds, except for copper, presents a relatively narrow distribution while cooper-oxygen bonds show two distinct peaks. This phenomenon can be attributed to the Jahn-Teller effect[52], commonly observed in octahedral complexes. Especially, in six-coordinate copper (II) complexes, the $d^9$ electron configuration of the ion gives three electrons in the two degenerate $e_g$ orbitals, leading to a doubly degenerate electronic ground state. To remove this orbital and electronic degeneracies to lower the total energy, distortion along one of the molecular fourfold axes (usually labelled as z axis) will occur. From Fig. 3(a) Cu-O distance distribution, it's evident to tell that the area of first peak is about double of the second peak, from which we can deduct the distortion around copper ion is extended along z axis. The two bonds along the elongated z axis is longer than the rest four horizontal short bonds. And after the adding of 6$^{th}$ dopant elements, which shown in Fig. 3 (b) - (f), the Cu-O bond have been substantially affected. The typical distribution of Jahn-Teller effect around copper ion is disrupted and displays different type of configurations. In case of Sc as the 6$^{th}$ dopant (J30), the peak corresponding to shorter Cu-O bonds disappears and the peak of longer Cu-O bonds broadens. This can be attributed to the electronic orbital of Sc ion, which is $3s^2 3p^6 d^1$. The outermost $d^1$ electron exactly make up for the non-fully-populated $d^9$ orbital of copper ions and hence reverse the [CuO$_6$] to normal state. And in the case of Sb as the 6$^{th}$ dopant (J31), things are different. The peak corresponding to shorter Cu-O bonds still exist but the peak of longer bonds disappear and evenly shift to a larger position. This indicates that the addition of Sb element results in being plundered of longer-bonded oxygen ion from copper ion. The electron orbital of Sb ion is $5s^3$, which means there will be 3 unpaired electrons left to influence other M-O bonds. And for oxygens around Sb ions, some longer Sb-O bonds cannot be ignored, which means a few oxygen ions around Sb ion are robbed away by other metal element. The situation happened in Sn-added composition (J34) is like J31. But for Sn ions, the outer most electron configuration is $5s^2$, which is one electron less than that in Sb ion. The 1NN M-O bonds distribution of Cr-added J35 is like J30 but with minute short bond peak. And for the case of J36, the 6$^{th}$ dopant element is Ge, which is at the same IV main-group as Sn but on top of it. Theoretically, the control exerted by Ge element on outer electrons is stronger than that of Sn.

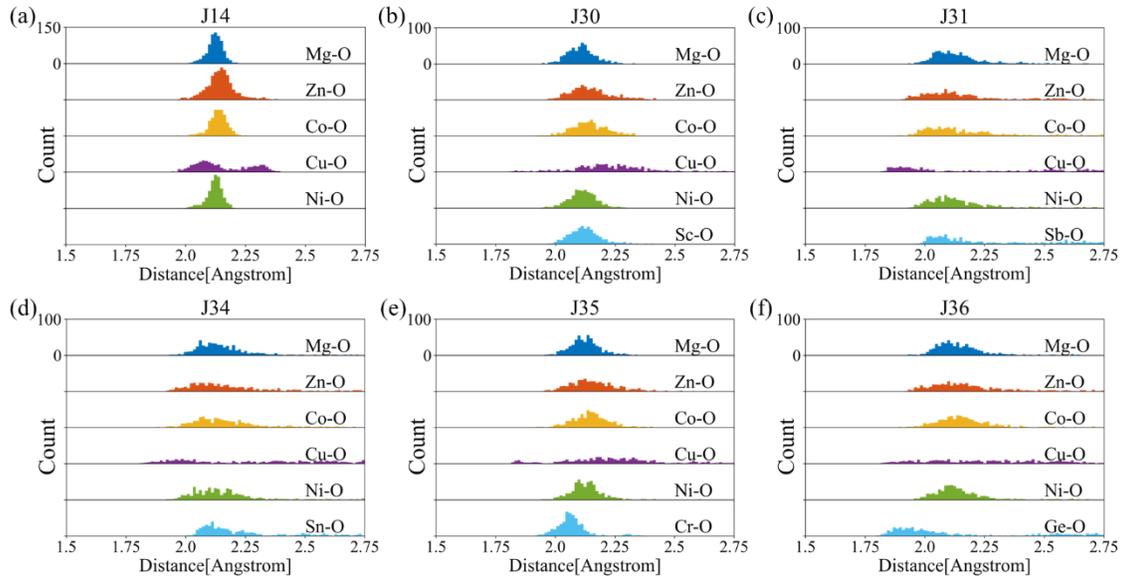

Figure 3 1NN Metal-Oxygen distance distribution histogram.

To quantitatively assess the lattice distortion in these six compositions, we calculated the average M-O bond distance and standard deviation, as shown in Fig. 4. The average M-O bond distances of Mg, Zn, Co, and Ni elements are quite close across different compositions. However, for the Cu and the sixth dopant element, the average M-O bond lengths exhibit significant variation among the different compositions. Additionally, the error bars in the graph represent the [$MO_6$] polyhedron distortion, which reflects the degree of deviation from the ideal geometry. The observed changes in the error bars are substantial. Specifically, in compositions J31, J34, and J36, the polyhedron distortion for all elements is pronounced. Conversely, in J30 and J35, only the Cu-O polyhedron exhibits significant distortion. This discrepancy primarily arises from the selection of the 6th dopant element. As discussed previously, elements like Sc and Cr have the ability to donate electrons to the copper ion, thereby reversing the Jahn-Teller effect. On the other hand, elements such as Sb, Sn, and Ge with unpaired electrons exert an influence on other M-O bonds, leading to the observed differences in distortion.

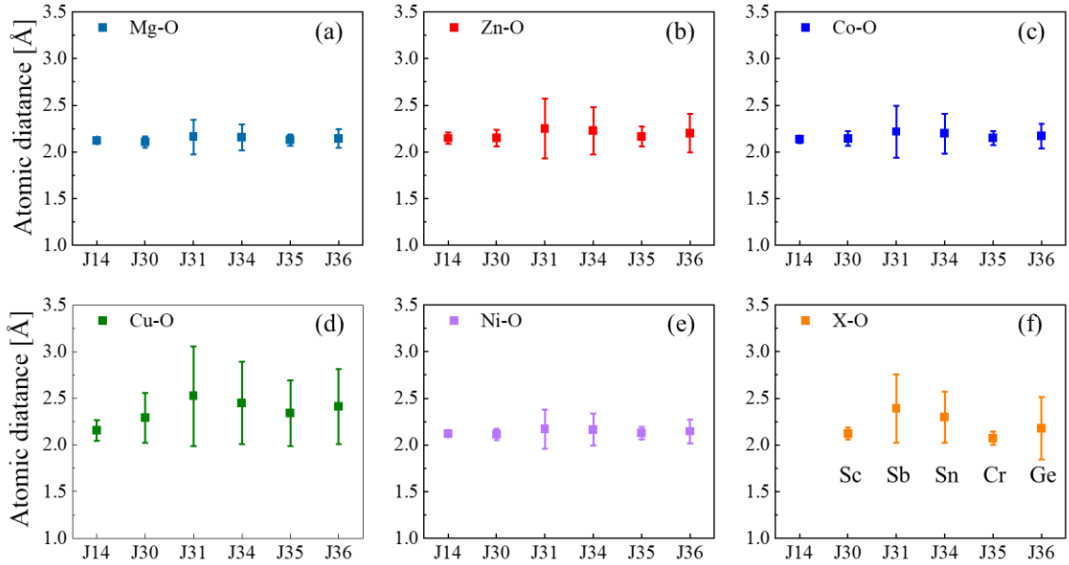

Figure 4 Average M-O distance and standard deviation in different compostions.

## 3.3 Unfolded spectra

After the relaxation of supercell, 2$^{nd}$ force constants of the supercell were obtained using finite difference method, and from which phonon unfolding spectra along X direction were calculated and shown in Fig. 5. The blue dash lines represent the results of virtual crystal approximation (VCA) method, which averages the effects of mass and force disorder. Basically, the configuration of phonon unfolding spectra aligns well with the curves of VCA results. For the upper longitudinal optical branches, the spectra center is slightly higher than that of VCA curve, which means the VCA underestimates the optical curves. But for acoustic branches, the results are opposite. Both the longitudinal and transvers phonon spectra of all the samples are slightly lower than that of VCA results. Compared to VCA, unfolded phonon dispersion is a spectrum and it has different distribution at different wave vector. And for phonon spectra of 5-dopant oxide J14, the spectra distribution is more compact and the intensity is concentrated in the red part. However, for other 6-dopant samples, the phonon spectra are more widely distributed. It's rare to see the red part in optical phonon branches and only the front and middle part of acoustic branches show concentrated red area. The value of spectra intensity come from the projection operator and have been normalized, which means, at any wavevector position, the integral of the spectra equals to the number of phonon branches. And this spectra intensity can be regarded as the probability of phonon dispersion appearance. What's more, its reported that these calculated phonon spectra can express the phonon scattering effects of mass and force disorder in high entropy materials, and this scattering effect far overwhelms the effect from phonon-phonon scattering[41, 44]. And most importantly, the linewidth of phonon spectra is reversely proportional to phonon scattering lifetime, which means it provides us the possibility to predict thermal conductivity. Consequently, the next step involves extracting the linewidth information from the phonon spectra and attempting to construct a prediction model for thermal conductivity. This will be further discussed in the subsequent subchapter.

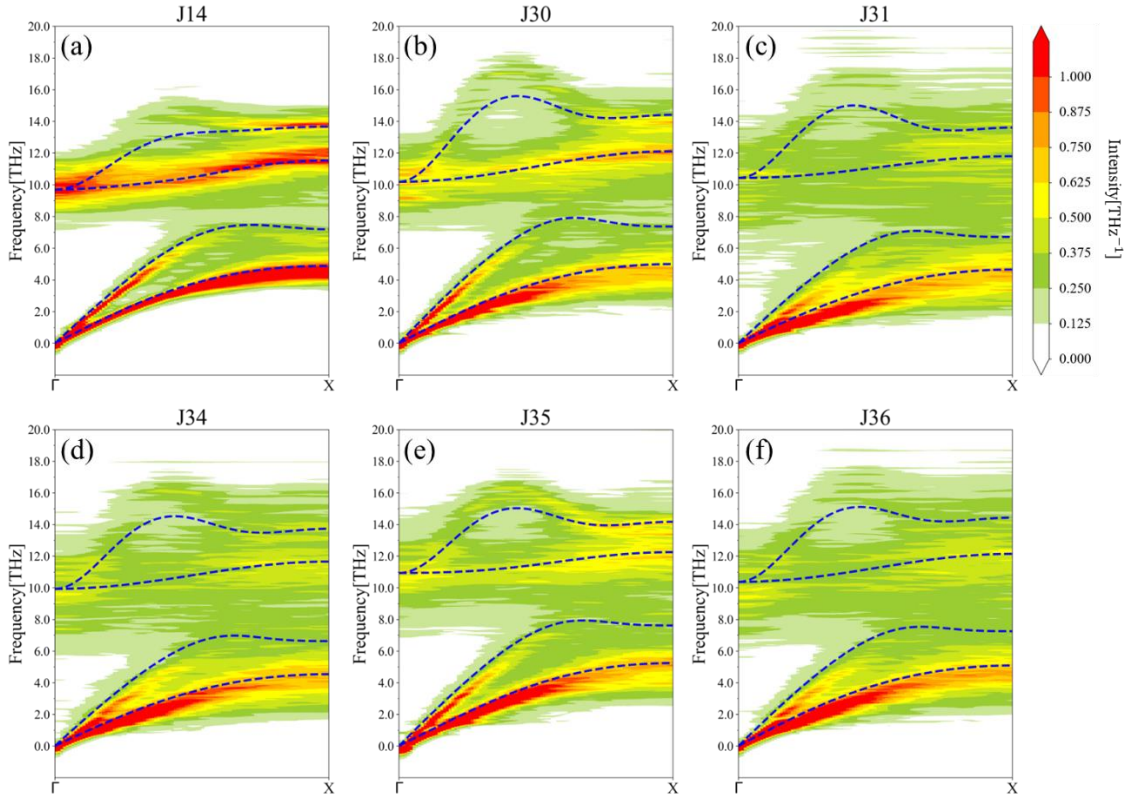

Figure 5 Unfolded phonon spectra of rocksalt high entropy oxides along X direction. The blue dash lines come from the results of VCA.

## 3.4 Thermal conductivity prediction

According to Boltzmann equation with relaxation time approximation, thermal conductivity can be expressed as:

$$\kappa = \frac{1}{NV_0} \sum_\lambda Cv_\lambda \cdot v_\lambda^2 \cdot \tau_\lambda \qquad (1)$$

where $N$ is the number of wave vectors, $V_0$ is the volume of primitive cell, $Cv$ is heat capacity, $v$ is the speed of sound (phonon group velocity), $\tau$ is phonon lifetime and $\lambda$ is grid wave vector point. This equation combines the contributions from heat capacity, speed of sound and phonon lifetime to determine thermal conductivity. Considering phonon lifetime in high entropy system can be regarded as inversely proportional to unfolded phonon spectra linewidth, our predicted thermal conductivity can be written as:

$$\Lambda = \frac{A}{NV_0} \sum_\lambda \frac{Cv_\lambda \cdot v_{\lambda}^2{}_x}{lw_\lambda} \qquad (2)$$

where $\Lambda$ is our predictor, $lw$ is phonon spectra linewidth, $A$ is adjustment factor to be fitted.

Fig. 6 is the phonon spectra linewidths of high entropy rocksalt oxides, which were extracted by applying Lorentzian function fitter. To ensure accuracy and account for randomness, five random supercells were constructed for each composition, followed by unfolding calculations. The linewidth curves presented in Fig. 6 represent the averaged results of these five repetitions, thereby eliminating calculation randomness. The graph clearly shows that the linewidth of J14 has the lowest profile while J31 has the highest profile, which means J14 has the longest phonon lifetime and J31 has the shortest. A notable difference in linewidth trends can be observed between the optical and acoustic branches along the X direction. By and large, the linewidths of optical branches display a negative relationship with the phonon dispersion curve, whereas the linewidths of acoustic branches show a positive relationship with the phonon dispersion. This phenomenon is also found in molecular dynamic research[53]. Considering the significant contribution of acoustic branches to thermal conductivity due to their larger phonon velocities and lifetimes, our primary focus should be on the linewidths of the acoustic branches. The acoustic linewidth difference of J14 with other five 6-dopant oxides is substantial. Firstly, both the longitudinal and transverse acoustic phonon linewidth of 6-dopant oxides are roughly double those of the 5-dopant J14. Secondly, the front half of the acoustic transverse phonon linewidth of J14 is flat and gradually increase in the second half zone. And for 6-dopant oxides, they exhibit a continuous increase in phonon spectra linewidths along the entire wavevector direction (X). This indicates that the addition of the sixth dopant element strengthen the scattering of acoustic transverse phonons in long wavelength range. And within the 6-dopant oxides, the differences primarily concentrate on the fluctuations in acoustic longitudinal phonon linewidths near the Gamma point. Notably, around the Gamma point, the linewidth curves of acoustic branches are not entirely smooth and may exhibit one or more bumps. Some studies [35, 42] regard this phonon-disorder scattering as Rayleigh scattering since the broadening of phonon branches is less in long wavelength zone than the short wavelength zone.

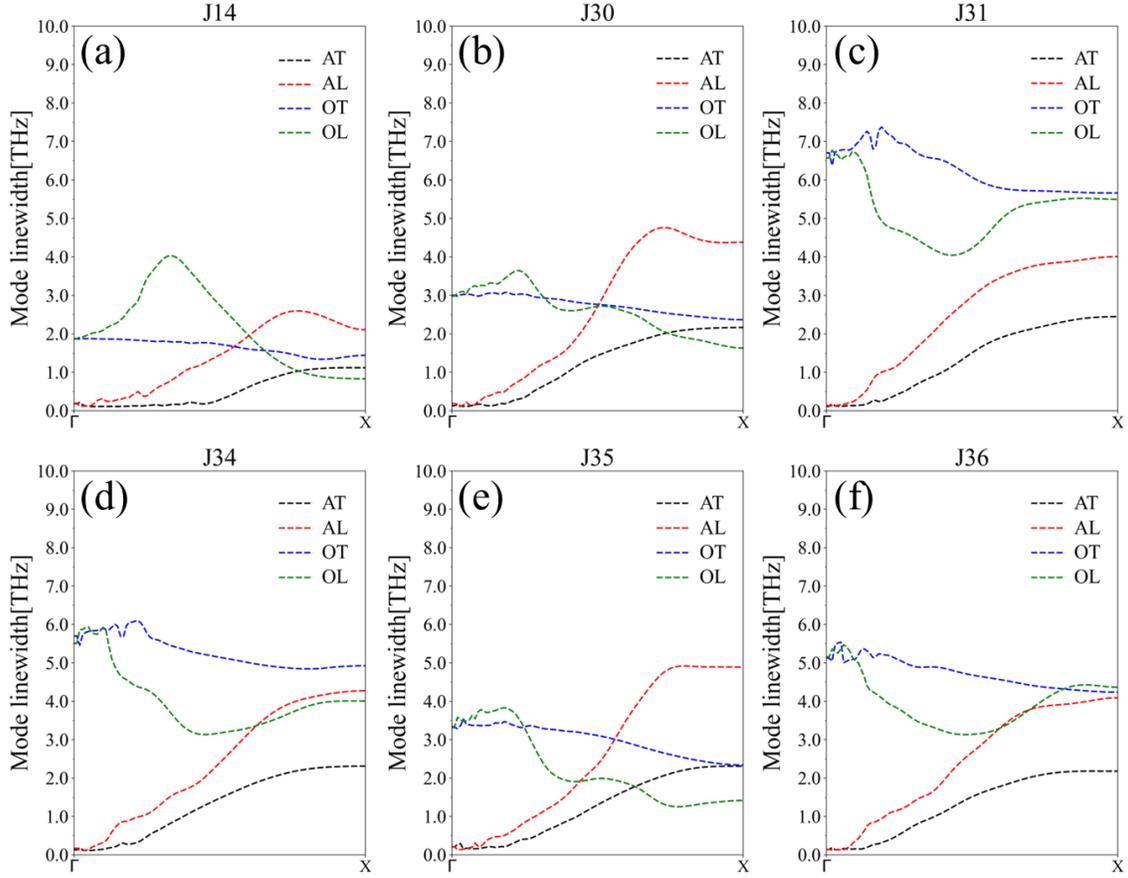

Figure 6 Phonon linewidths of rocksalt high entropy oxides.

Furthermore, the other two factors, the mode heat capacity (room temperature) and phonon velocity, can be approximated directly using the *Phonopy* Package based on the virtual crystal approximation (VCA) model. And for the other two factors, the mode heat capacity (room temperature) and phonon velocity, they can be directly approximated by Phonopy Package based on VCA model. In this paper, we will not discuss the parameter *A*, which may be dependent on the crystal structure and should remain the same within a given system. Therefore, we assume its value to be 1 for simplicity.

The prediction results along the reciprocal direction from Γ to X (corresponding to the [001] direction in real space) are presented in Fig. 7, alongside the experimental results for comparison. After adjusting the scale of the y-coordinate, we observe a close match between our predicted values and the experimental results. This alignment suggests the validity of our assumptions and the feasibility of our prediction approach. However, for the J30 composition, a slight discrepancy can be observed between the experimental and predicted values. This difference can be attributed to the formation of vacancies in the experimental samples. The addition of the Sc element may result in a higher vacancy concentration compared to the other elements. These vacancies disrupt the crystal's continuity, distort the lattice, and substantially reduce the thermal conductivity.

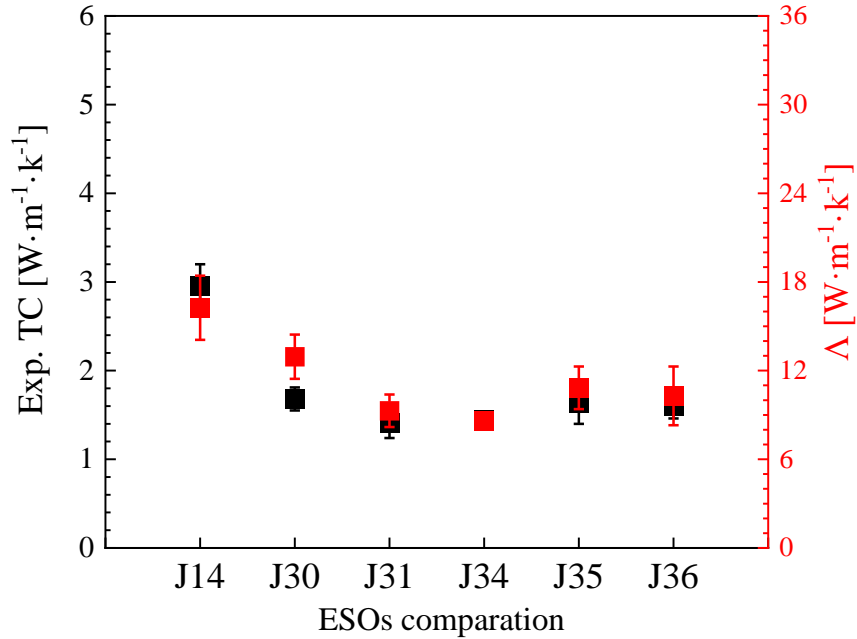

Figure 7 Comparation between experimental results and calculated indicater of thermal conductivities.

Indeed, it is evident that the disorder-phonon scattering in our high entropy system arises from both force and mass disorder. However, considering the compositional selection in our study, the only notable difference lies in the sixth dopant element, which implies that the mass differences among the compositions are not prominent. Consequently, we can focus our investigation on the force disorder within the high entropy system. Fig. 6 depicts the distribution of force constant $\Phi_{xy}$ for the first-nearest-neighbor (1NN) metal-oxygen pair. Compared to $\Phi_{xx}$ distribution which is distributedly disorganized, the distribution of $\Phi_{xy}$ exhibits a more regular pattern that can be analyzed. We employed a Lorentzian function to fit these $\Phi_{xy}$ distributions, as shown in Fig.8. The FWHM can denote the wide of the distribution and to some degree represents the force disorder of shear stress $\Phi_{xy}$. The FWHM of J14 is 0.0729, which is roughly half the value of the rest 6-dopant oxides, which means that addition of the sixth dopant element largely broaden the shear stress distributin of 1NN. And this broadening of stress force distribution should be responsibile for the linewidths change of acoustic transverse branch in Fig 6, from flat curve of

J14 to gradually increasing curve of other 6-dopant oxides around the gamma point. Moreover, if compare the FWHM of $\Phi_{xy}$ with thermal conductivities of all the samples, it's found that their correlation is close to inversely proportion. This is a very interesting relationship and need further study.

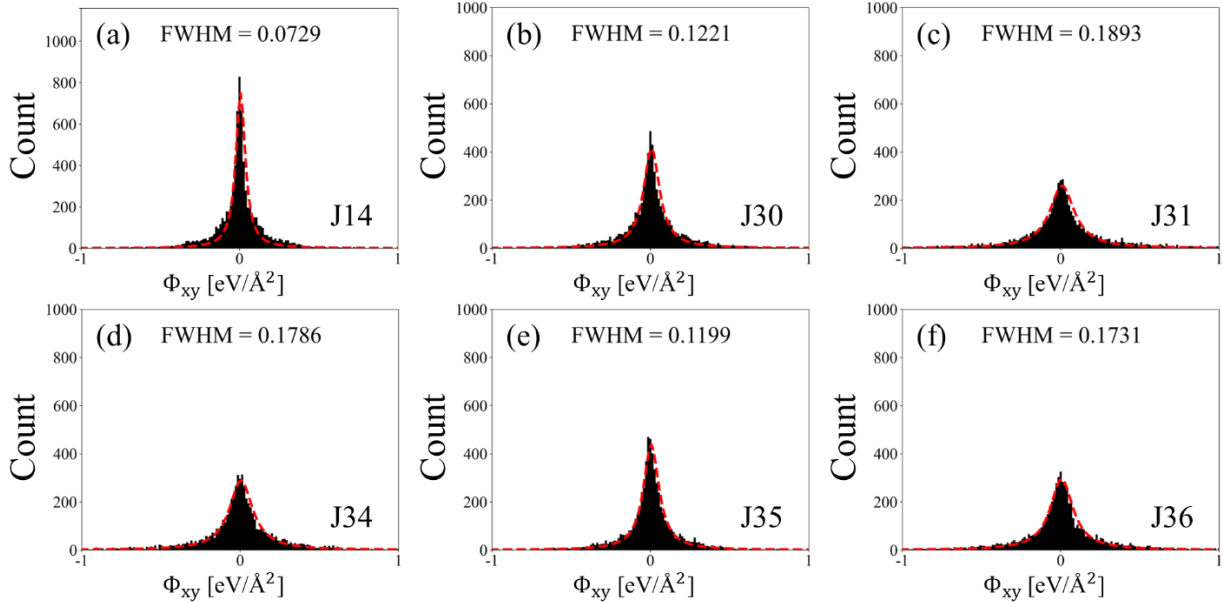

Figure 8 Force constant $\Phi_{xy}$ distribution of first-nearest-neighbor X-O (X denote metal atoms) and lorentzian fitting.

## 4. Summary

In this research, we have proposed a novel method to estimate the thermal conductivity of high entropy ceramics based on the phonon-unfolding method. Specifically, we apply this method to rock salt high entropy oxides and the results demonstrate a strong agreement between our thermal conductivity predictions and experimental results. Additionally, we have analyzed the disorder mechanism of high entropy rock salt oxides in relation to the 6[th] dopant elements. It is observed that the presence of the 6[th] dopants element can affect the Jahn-Teller effects around the copper ions, resulting in variations of overall lattice distortion, and consequently causes the force disorder.

The unfolded phonon spectra calculations assume that the atoms remain at their ideal positions while considering disordered force constants of relaxed lattice structure, which slightly deviates from the actual conditions in the materials. Nonetheless, this method works well for predicting thermal conductivity in high entropy ceramics. We must admit that there are still certain unknown factors that require further investigation to fully understand the mechanism of thermal conduction in high entropy ceramics. The phonon-disorder scattering mechanism and the quantification of mass and force disorder are areas that warrant continued study.

# Acknowledgements

The author acknowledges the usage of the Cedar HPC cluster provided by Digital Research Alliance of Canada for providing computing resources. This work was supported by National Science and Engineering Research Council of Canada (NSERC).